\documentclass{elsart3p}
\usepackage{natbib}
\usepackage{graphics}
\usepackage{graphicx}
\usepackage{epsfig}

\usepackage{amssymb}

\journal{Nucl. Instr. Meth. A}

\begin{document}

\begin{frontmatter}

\title{The Proto Type of Shanghai Laser Electron Gamma Source at 100 MeV LINAC}

\author[SINAP,ZJFC]{J.G. Chen\corauthref{cor1}},\corauth[cor1]{Corresponding author.}\ead{chenjg@sinap.ac.cn}
\author[SINAP]{W. Xu},
\author[SINAP,GSCAS]{W. Guo},
\author[SINAP]{Y.G. Ma},
\author[SINAP]{X.Z. Cai},
\author[SINAP]{H.W. Wang},
\author[SINAP]{G.C. Lu},
\author[SINAP,GSCAS]{Y. Xu},
\author[SINAP,GSCAS]{C.B. Wang},
\author[SINAP]{Q.Y. Pan},
\author[SINAP]{R.Y. Yuan},
\author[SINAP]{J.Q. Xu},
\author[SINAP]{Z.Y. Wei},
\author[SINAP]{Z. Yan},
\author[SINAP]{W.Q. Shen}

\address[SINAP]{Shanghai Institute of Applied Physics, Chinese
Academy of Sciences, Shanghai 201800, P.R.China}
\address[ZJFC]{College of Science,
Zhejiang Forestry University, Hangzhou 311300, P.R.China}
\address[GSCAS]{Graduate School of the
Chinese Academy of Sciences, Beijing 100039, P.R.China}

\begin{abstract}
The design for the proto type of the Shanghai Laser Electron Gamma
Source (SLEGS) at the Shanghai Synchrotron Radiation Facility
(SSRF) is introduced. Some detailed descriptions for design of
related instruments are provided. The proto type can produce X-ray
with energy of 10 keV order. A description of the kinematics of
Compton backscattering mechanism and the related simulation
results are presented and discussed. The backgrounds from dipole
magnet and bremsstrahlung are estimated and the signal-noise ratio
is also given.
\end{abstract}

\end{frontmatter}

\section{Introduction}
With the improvement of accelerator technology, various photon
sources around the world are set up based on laser Compton
scattering process
\cite{CAS75,MAT77,FED80A,BAB91,DAN00,PAR01,LIT97,OHK06,FUJ03,AOK04,LI04},
which possesses several advantages, such as rather flat energy
distribution with small spreading and high linear- or
circular-polarization \cite{SAL95}.

The Shanghai Synchrotron Radiation Facility (SSRF) is designed for
the advanced third generation moderate energy synchrotron
radiation facility \cite{ZHA04}. The synchrotron radiation ray
produced from the SSRF covers a broad wave band from far infrared
to hard X-ray. It has a low emittance of moderate energy
electrons. The Shanghai Laser Electron Gamma Source (SLEGS) will
be built at one of the straight sections of the SSRF. Gamma-rays
with energy up to 22 MeV will be produced by Compton
backscattering of far infrared laser photons on the 3.5 GeV
electrons circulating in the synchrotron ring. The SLEGS will be
used for research in astrophysics, nuclear physics and
applications for material sciences, etc. So the proposed
construction of SLEGS will be able to provide us uncommon
opportunities to probe deeply sub-nuclear structure and expand our
research methods.

At present, The SLEGS is at the stage of construction of its proto
type. The proto type will be set up at the terminal of a 100 MeV
LINAC. X-rays at the energy of 10 keV order can be produced from
the proto type. Due to the limited capabilities of the 100 MeV
LINAC and the laser, the brightness of X-ray produced from the
proto type will be low. However, this is not so serious, because
the primary object for the construction of the proto type is to
pave the way for the future setup of the SLEGS.

\section{Basic physics}

\subsection{ Energy of the Compton backscattered photon}

\begin{figure}
\includegraphics[scale=0.55,bb=-30 -15 138 180]{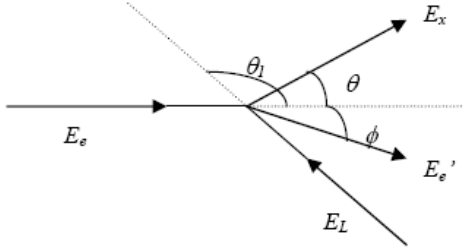}
\caption{The Compton scattering process} \label{scattering}
\end{figure}

The scattering process between the CO$_2$ laser photons and the
electrons from  100 MeV LINAC is shown schematically in Fig.
\ref{scattering}. Here $E_e$ and $E_{e}^{'}$ are the kinetic
energies of incident and scattered electron and $E_L$ and $E_x$
are the energies of incident and scattered laser photon in the
laboratory frame. $\theta_1$ (If $\theta_1$ equals $\pi$, this
scattering can be simplified to strict backscattering.) is the
laser incident angle relative to the electron beam direction in
the laboratory frame. After backscattering, the X-ray emerges in
the laboratory frame at a small angle $\theta$ relative to the
electron beam direction and the electron emerges at the angle
$\phi$.

The theoretical basic of Compton backscattering mechanism was
described in Refs. \cite{FED80B,SAN82,BLU92}. For a better
understanding of this process, some important relationships will
be displayed as below. One can obtain the energy of scattered
photon with the help of equations for energy and momentum
conservation:
\begin{equation}
E_{x}=\frac{(1-\beta\cos{\theta_1})E_{L}}
{1-\beta\cos\theta+(1-\cos{(\theta-\theta_1)})E_{L}/E_{e}}
\label{energy1}
\end{equation}
where $\beta$ is the ratio of the electron and light velocities.
Considering the electron energies above 50 MeV
($\beta\rightarrow$1), and a small scattering angle ($\theta<$10
mrad), formula (2) can be simplified to
\begin{equation}
E_{x}=\frac{4\gamma^{2}E_{L}} {1+4\gamma
E_{L}/m_{0}c^2+\gamma^{2}\theta^{2}}
\end{equation}
given by Federici et al. \cite{FED80B}, Sandorifi et al.
\cite{SAN82} and Blumberg \cite{BLU92}, where $\gamma$ is the
ratio of the total electron energy and its energy at rest:
$\gamma=1/\sqrt{1-\beta^{2}}$ and $m_0$ is the electron mass at
rest.

\subsection{Compton scattering cross sections}

The differential Compton scattering cross section for energetic
electrons in the laboratory frame can be obtained starting with
its form in the ER frame, in which the cross section was derived
by Klein-Nishina \cite{KLE29}:
\begin{equation}
\frac{d\sigma}{d\Omega}=\frac{r^{2}_{0}}{2}R^{2}
\bigg(R+\frac{1}{R}-1+\cos^{2} \theta_{ER}\bigg)
\label{KleinNishina-ER}
\end{equation}
where $r_0$, $R$ and $\theta_{ER}$ are the classical electron
radius, the ratio between the energies of scattered and incident
photon, and the scattering angle in the ER frame, respectively. It
can be obtained from Eq. (2) considering the electron to be at
rest, i.e. $\beta=0$. In this case, the ER and laboratory frame
are equivalent. The result is:
\begin{equation}
R=\frac{E_{x}}{E_{L}}=\frac{1}{1+(E_{L}/{m_0c^2})(1+\cos\theta_{ER})}\label{ratio}
\end{equation}
Then one can obtain the Compton scattering cross section for
energetic electrons in laboratory frame via Lorenz transformation:
\begin{eqnarray}
\nonumber \frac{d\sigma}{\sin\theta d\theta} & = & \pi r^{2}_{0}
\frac{1-\beta^{2}}{(1-\beta \cos \theta)^{2}}\times \\
& & R^{2} \bigg(R+\frac{1}{R}-1+\cos^{2} \theta_{ER}\bigg)
\label{KleinNishina-Lab}
\end{eqnarray}

Energy differential cross-section of Compton backscattering  X
photons can be derived from formula (\ref{energy1}) and
(\ref{KleinNishina-Lab}):
\begin{eqnarray}
\nonumber \frac{d\sigma}{d E_x} & = & \pi r^{2}_{0}
\frac{1-\beta^{2}}{(1-\beta \cos \theta)^{2}}
\frac{(1+\beta)E_{L}}{E_{x}^{2}(\beta - E_{L}/E_{0})} \times\\
& & R^{2} \bigg(R+\frac{1}{R}-1+\cos^{2} \theta_{ER}\bigg)
\end{eqnarray}

When injected laser beam is totally polarized, circular and linear
polarization of BCS photons are presented as
\begin{equation}
\left\{\begin{array}{l}P_C=\frac{-\cos \theta _{ER}}{R+1/R-1+\cos
^{2} \theta _{ER}}
\bigg(R+\frac{1}{R}\bigg)\\
P_L=\frac{1}{2}\frac{(1+\cos \theta _{ER})^{2}} {R+1/R-1+\cos ^{2}
\theta _{ER}} \label{Polarization}
\end{array}\right.
\end{equation}

\section{The proto type of the SLEGS facility}
The proto type of the SLEGS facility consists mainly of 100 MeV
LINAC, CO$_{2}$ laser optical system, target chamber, detector
system and data acquisition system, etc. They will be introduced
in detail in the following.

\subsection{The 100 MeV LINAC}

\begin{figure*}[htpb]
\includegraphics[scale=0.75,bb=600 305 0 532]{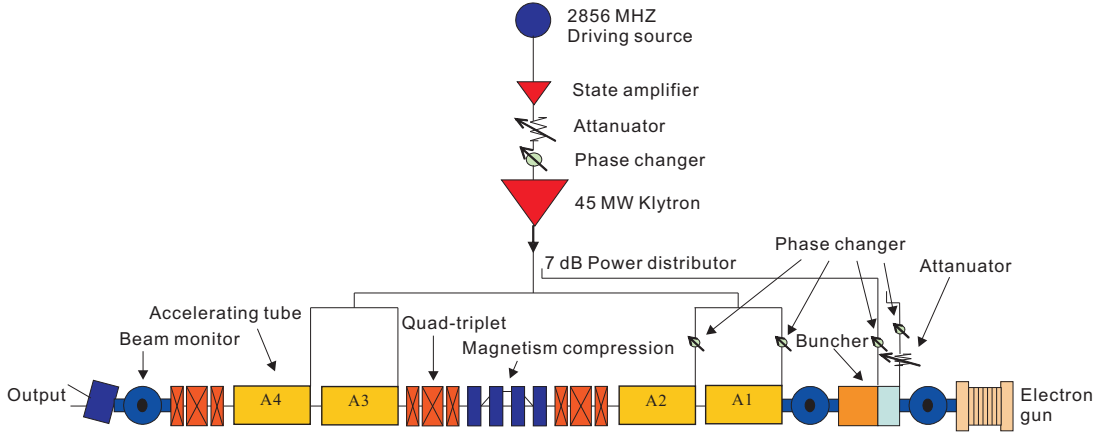}
\centering  \caption{The schematic illustration for the 100 MeV
LINAC} \label{linac}
\end{figure*}

The schematic illustration for the 100 MeV LINAC is shown in Fig.
\ref{linac}. It is mainly composed by an electron gun, an
pre-buncher and a buncher, four accelerating tubes, a focus coil,
a klystron, a modulator, a microwave driving source, a phase
control system, a vacuum system, a water-cooling system, a beam
flux diagnosis system and a control system, etc. The total length
is about 20 meters. Each accelerating tube may make the electronic
energy to increase approximately 27.5 MeV. About 7 MW power from
klystron is input to the buncher via the wave guide directional
coupler. The klystron outputs the other microwave power to the
four accelerating tubes via the wave guide merit minute, each
accelerating tube power accepts about 7 MW.

\subsection{Optical system}
\begin{figure}[htpb]
\includegraphics[scale=0.75,bb=600 317 0 744]{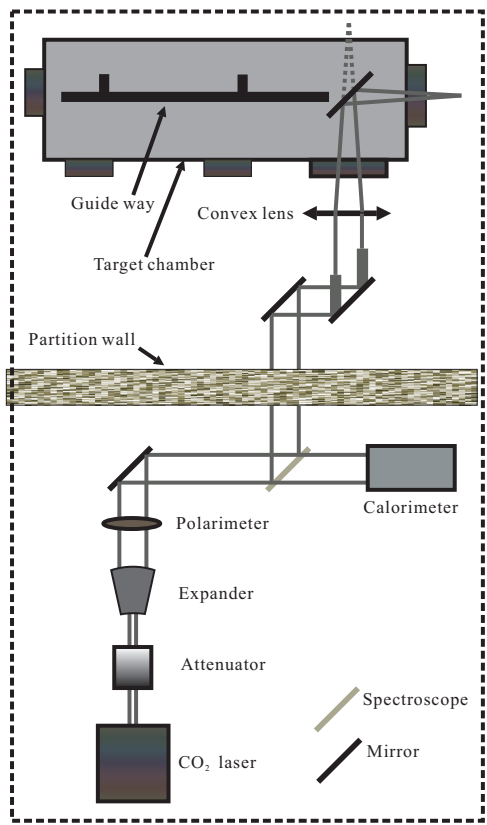}
\centering  \caption{The schematic view of CO$_2$ laser optical
system.} \label{optical}
\end{figure}

Fig. \ref{optical} display a schematic view for optical system of
proto type. First, the CO2 laser will pass through an energy
attenuator and then be expanded. A suitable adjustment of the
positions of the two lenses will make the laser light be parallel.
Then, the parallel light will be polarized linearly or circularly
via the polarizing disc. Finally, the laser will be reflected and
then be guided to the 100 MeV LINAC hall for the preparation of
collision with electron beam. In the 100 MeV LINAC hall, the laser
will be adjusted again, to make it pass vertically through target
window. The details for the adjustment of laser and electron beam
will be provide in following.

\subsection{Target chamber}
\begin{figure}[htpb]
\includegraphics[scale=0.55,bb=210 410 345 570]{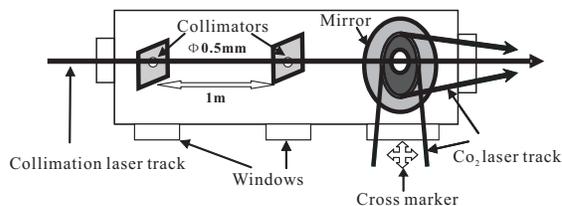}
\centering \caption{ The collimation of CO$_2$ laser and electron
beam. } \label{collimation}
\end{figure}

In order to reduce the loss of the CO$_2$ laser power, the laser
should transmit in a vacuum chamber. According to the present
situation of 100 MeV LINAC, we plan to develop a vacuum target
chamber, which is used for the entrance of laser. The schematic
view for target chamber is displayed in Fig. \ref{collimation}. As
well known, the collimation between the CO$_2$ laser and the
electron beam is very important, because it decides the energy
region and the flux of X-ray. Fig. \ref{collimation}. displays
step by step the aim between the CO$_2$ laser and the electron
beam. There are two flakes filmed fluorescence and a mirror with a
hole at its center for the transmission of laser. First, one
should initialize the aim via marking the track of electron beam
on the two flakes. Second, another laser used for collimation will
be imported and adjusted to be coaxial with the electron beam.
Then the CO$_2$ laser should be adjust to be coaxial with
collimated laser. The cross marker should be used for the
benchmark for the entrance of the CO$_2$ laser. Third, to readjust
the CO$_2$ laser after the CO$_2$ laser being fixed at the LINAC.
Finally, to readjust the electron beam to make it be coaxial with
the CO$_2$ laser.

\subsection{Detector system}
The X-ray from Compton back scattering performed at the proto type
has an energy range of 5 keV$\thicksim$18 keV and the
corresponding wave length range is about $2.5\times10^{-8}$
cm$\thicksim$$0.7\times10^{-8}$ cm (2.5$\thicksim$0.7 \AA).
Therefore a HPGe detector or a Si (Li) detector is best suitable,
because their survey energy scope is about 1 KeV$\thicksim$30 KeV.
The efficient area is larger than 12.5 mm$^2$ and the energy
resolution (FWHM) is smaller than 160 eV at 5.9 KeV. The related
parameters for Si (Li) detector are listed in Table \ref{Sili}.

\begin{table}[htbp]
\centering\caption{Basic parameters for Si (Li)
detector.}\tabcolsep 0.7cm
\begin{tabular}{l l}
\hline \hline

Diameter & 16 mm \\
Efficient area & 200 mm \\
Energy resolution &  220 eV \\
Thickness of Be window &  50 $\mu$m  \\
Energy range & 1 $\thicksim$ 30 KeV \\

\hline\hline
\end{tabular}
\label{Sili}
\end{table}

\subsection{DAQ}
Here the data acquisition refers mainly to the acquisition and
processing for data from the detector terminal. So data
acquisition system includes standard CAMAC and the VME data bus
system, which are adopted frequently in nuclear physics as well as
in high energy physics experiment.

\begin{figure}[htpb]
\includegraphics[scale=0.8,bb=220 35 10 322]{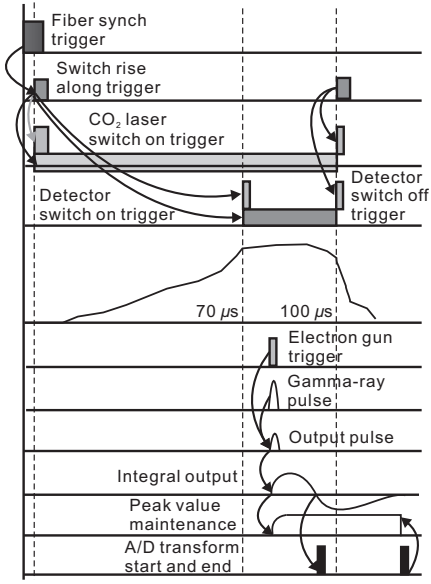}
\centering  \caption{The synch control for data acquisition
system} \label{daq}
\end{figure}

Fig. \ref{daq} displays the control illustration for DAQ. A synch
trigger signal is provided by the optical fiber of the SSRF synch
system. Then the light signal is transformed by the
photoelectricity coupled apparatus into the switch quantity
electrical signal, which is used to trigger the CO$_2$ laser. When
the laser output power approaches peak value after an exponential
increase with 70 $\mu$s, the CO$_2$ laser synch system will
trigger the permission logical gate of the gamma detector. The
detecting time is 30 $\mu$s.

When gamma rays are detected, they will be transformed into
electricity pulse, which is proportional to the energy deposits of
gamma rays in the detector. This pulse will be reshaped and
enlarged with a maintenance of peak value. Then it passes through
zero fixed time to output the A/D transformation logic pulse.
After the A/D converter transforms being finished, a logic signal
will be produced to end the maintenance of peak value.

\section{Some simulated results and discussion}

Various useful parameters for 100 MeV LINAC and for CO$_2$ laser
are listed in Table \ref{parameter}. All the required quantities
are calculated based on the above formulas and the parameters
listed in Table \ref{parameter}.

\begin{table}[htbp]
\centering\caption{Basic parameters for 100 MeV LINAC and for
CO$_2$ laser.}\tabcolsep 0.2cm
\begin{tabular}{l l }
\hline \hline
\multicolumn{2}{l}{\it{\textbf{100 MeV LINAC}}}    \\
\hspace{5mm}Electron energy & \hspace{5mm}100 MeV \\
\hspace{5mm}Pulse width & $\hspace{5mm}\sim$324 ns (FHWM) \\
\hspace{5mm}Normalized emittance &\hspace{5mm}107 (x)$\sim$137 (y) $\mu$mrad \\
\hspace{5mm}Energy resolution &\hspace{5mm}$<$0.9$\%$ \\
\hspace{5mm}Mean current & $\hspace{5mm}\sim$100 mA  \\
\multicolumn{2}{l}{\it{\textbf{CO$_2$ laser}}}\\
\hspace{5mm}Wave length & \hspace{5mm}10.6$\pm$2 $\mu$m \\
\hspace{5mm}Peak power & \hspace{5mm}0.25$\sim$1.5 kW \\
\hspace{5mm}Rise/Fall time &\hspace{5mm}$<$90 $\mu$s \\
\hspace{5mm}Stabilization of power & \hspace{5mm}$<$8$\%$ \\
\hline\hline
\end{tabular}
\label{parameter}
\end{table}

\begin{figure}[htpb]
\includegraphics[scale=2.1,bb=10 16 0 203]{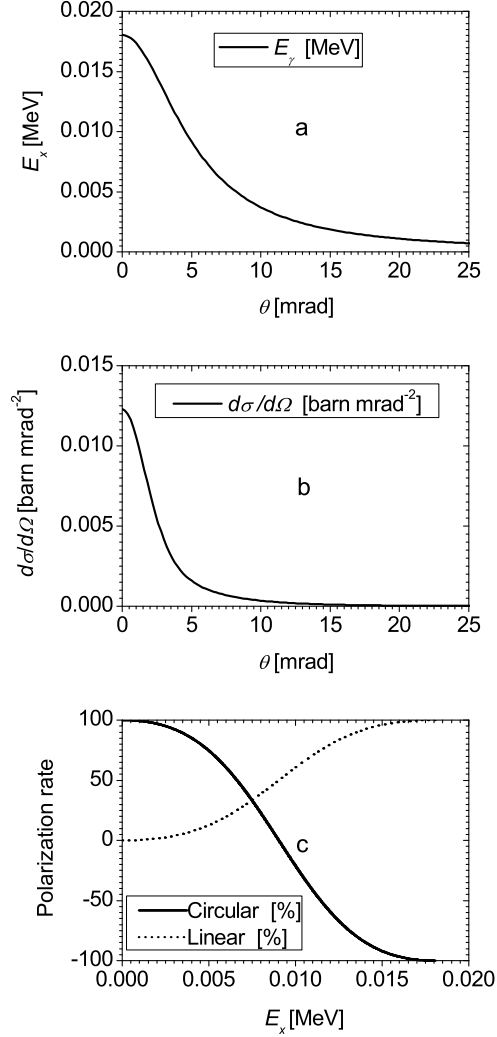}
\caption{The Compton scattering energy distribution (a),
differential cross section (b) and polarization rate (c). }
\label{ener-section-pola}
\end{figure}

Fig. \ref{ener-section-pola}a shows the relation between the
energy of Compton backscattering photons and the scattered angle.
The scattered angles of Compton backscattering photons above 10
keV are less than 4.6 mrad which corresponds to a beam size of 3.6
cm at the distance of 4 m from the collision point. Fig.
\ref{ener-section-pola}b displays the differential cross section
of Compton backscattering photons. It shows that the differential
cross section increases sharply when the scattered angle is less
than 5 mrad, which is consistent with interesting energy region
above. The polarization of Compton backscattering photons is given
in Fig. \ref{ener-section-pola}c. It can be seen that if laser
light is 100\% polarized, a Compton backscattered photon is highly
polarized at the maximum energy. The polarization drops as the
photon energy decreases as shown in Fig. \ref{ener-section-pola}c.
However, the energy of laser photons is easily changed, so the
polarization can remain be maintained reasonably high in the
energy region of interest.

\begin{figure}[htpb]
\includegraphics[scale=0.8,bb=190 15 100 210]{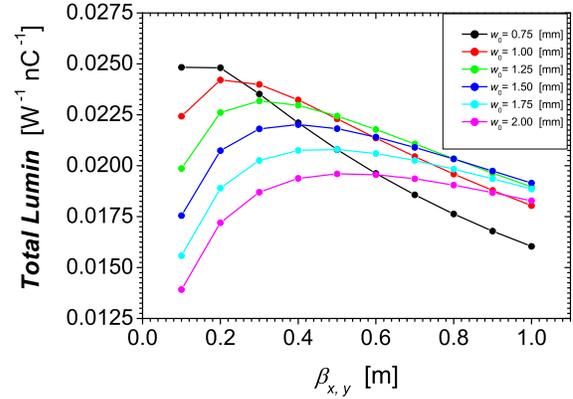}
\centering  \caption{Total X-ray flux as a function of TWISS
function.} \label{lumin}
\end{figure}

Differential luminous flux of Compton backscattering photon as a
funtion of interaction length at different laser waist widths is
shown in the upper part of Fig. \ref{lumin}. Here the electron
beam is assumed to be zero emittance and its transverse widths are
200 $\mu$m for $x$ direction and 100 $\mu$m for $y$ direction.
These two quantities will be perhaps changed a little in the real
operation of the LINAC. In addition, the electron beam and the
laser are assumed to be co-axis and co-focus. It is seen obviously
that differential flux distributions along interaction length are
different at different laser waist widths. Differential flux
changes sharply with the transform of interaction position if
laser waist width is very small but becomes insensitive when waist
width of laser is large enough.

\section{Estimation of background}
\begin{figure}[htpb]
\includegraphics[scale=1.8,bb=15 15 180 108]{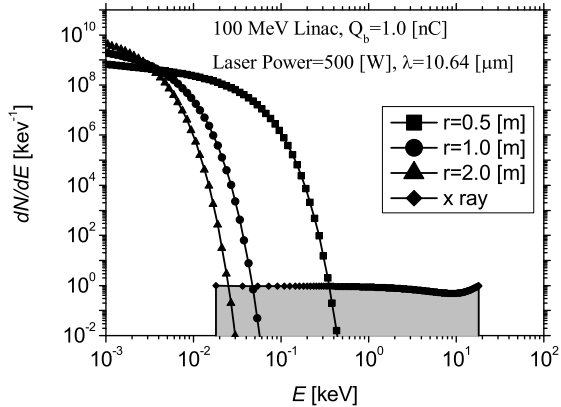}
\centering \caption{ The estimated background from synchrotron
radiation. } \label{synch}
\end{figure}

\begin{figure}[htpb]
\includegraphics[scale=1.5,bb=16 15 220 120]{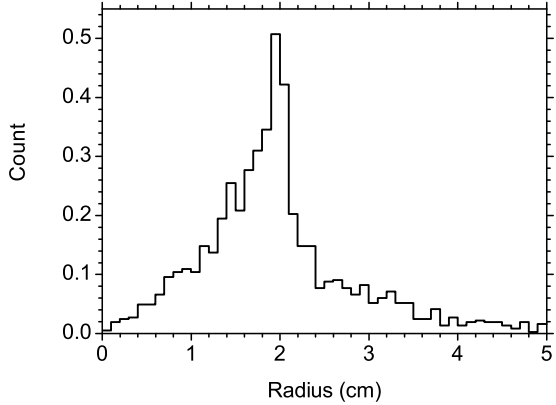}
\centering \caption{ The estimated background from bremsstrahlung.
} \label{brem}
\end{figure}

The backgrounds from dipole magnet and bremsstrahlung for the
proto type of SLEGS are estimated and the corresponding results
are displayed in Figs. \ref{synch} and \ref{brem}, respectively.

Since the radius of dipole magnet is not decided yet, three
candidate radii are assumed in the calculations, which are
performed using the program Xop 2.1 developed by European
Synchrotron Radiation Facility (ESRF) group. The fluxes of photons
from dipole magnet at three radii are much higher than that of X
photons. But the energies of background are very small (less than
1 keV), which is lower than the energy of X photons of interest.
So the background from dipole magnet will not cover up the X
photons of interested energy region.

\section{Conclusion}

In conclusion, the design of proto type of SLEGS at SSRF is
introduced. The related simulation results are presented and
discussed. The backgrounds from dipole magnet and bremsstrahlung
are estimated, respectively. From above calculations, the ideal
obtained flux of X-ray is about 0.076
s$^{-1}$$\cdot$W$^{-1}$$\cdot$nC$^{-1}$. However considering the
limited reception of detector and other X-ray loss, the real flux
of X-ray will be larger than 0.002
s$^{-1}$$\cdot$W$^{-1}$$\cdot$nC$^{-1}$. If a K-500 CO$_2$ laser
is selected, its maximum power is about 1500 W. Based on the above
estimation for background, the signal-noise ratio for the proto
type will be larger than 30.

\section{Acknowledgments} This work was supported by the
Innovation Program of Science $\&$ Technology of Chinese Academy
of Sciences under Contract No. KJCX2-SW-N13, by the National
Natural Science Foundation of China under Contract No. 10475108,
and by the hundred talent project of SINAP.

\end{document}